\documentclass[11pt,reqno]{amsart}

\usepackage[margin=1in]{geometry}

\usepackage{multicol}
\usepackage{hyperref}
\usepackage{amssymb,multicol,subfigure,graphicx}
\usepackage{wrapfig}
\usepackage{amscd,amssymb,latexsym,subfigure,verbatim,hyperref,epsfig,xypic}

\usepackage{amssymb}
\usepackage{epsfig,color}
\numberwithin{equation}{section}

\begin{document}
\baselineskip=15pt

\title[]{ Sonia Kovalevsky Days: \\ the potential to inspire \vspace*{-3mm}}

\author[L. P. Schaposnik]{Laura P. Schaposnik}
\author[J. Unwin]{James Unwin
\vspace*{-8mm}}


\subjclass[]{}

\date{}

\maketitle

    \begin{multicols}{2}

Sonia Kovalevsky overcame adversity to become the first woman to receive a PhD in mathematics and has since become a role model for young women interested in math and science. The Association of Women in Mathematics has organized and sponsored Sonia Kovalevsky ({\it SK}) Days for Girls  for almost three decades. These one day events are held at colleges and universities throughout the country and  consist of a program of workshops, talks, and problem solving sessions for female  students aged 7-17. The activities of the SK Days are designed to encourage young women to continue their studies in mathematics, and to help them learn about  educational  possibilities.


%

With some guidance, organizing these events can be done very smoothly, and the whole experience can be extremely rewarding both for the students, as well as for the organisers and volunteer helpers. Indeed, witnessing the enthusiasm of the students  for learning new mathematics is priceless. In what follows we would like to give some advice on how to make these events successful, and encourage colleagues to host similar series of events at their institutions.

     Most of what we will explain here is based on our experience running the SK Days  at the University of Illinois at Chicago. A website for the Sonia Kovalevsky Days at UIC, which we have ran annually since 2015, is hosted at: 
     
 \begin{center}     \href{http://schapos.people.uic.edu/Sonia.html}{\textcolor{blue}{http://schapos.people.uic.edu/Sonia.html}} \end{center}
 
 \noindent The site includes an educational paper containing material for running different outreach events for young students. We ourselves  learned a lot about how to run successful events from Michelle Delcourt and her volunteers during our time together at UIUC.
  
\noindent{\bf  The structure of the day.} After a short registration period during which parents sign any relevant forms, our events begin with one of the volunteers giving
 an introduction to the AWM for the students and accompanying parents and teachers, and a brief presentation on the life and achievements of Sonia Kovalevsky. The presentation focuses in particular on highlighting   the remarkable obstacles that Kovalevsky was able to overcome during her life. The students are then presented with some icebreaker activities which allow them to learn each other's names and  get comfortable in the setting.  

After the introduction, participants are separated into groups to do the activities of the day in smaller classrooms -- experience has shown that groups of about 12-15 people are best, with one volunteer every 5 or 6 students. Hence, depending on the number of girls registered, one can made 1, 2 or 3 parallel sessions: 
\begin{center}
    \includegraphics[width=0.5\textwidth]{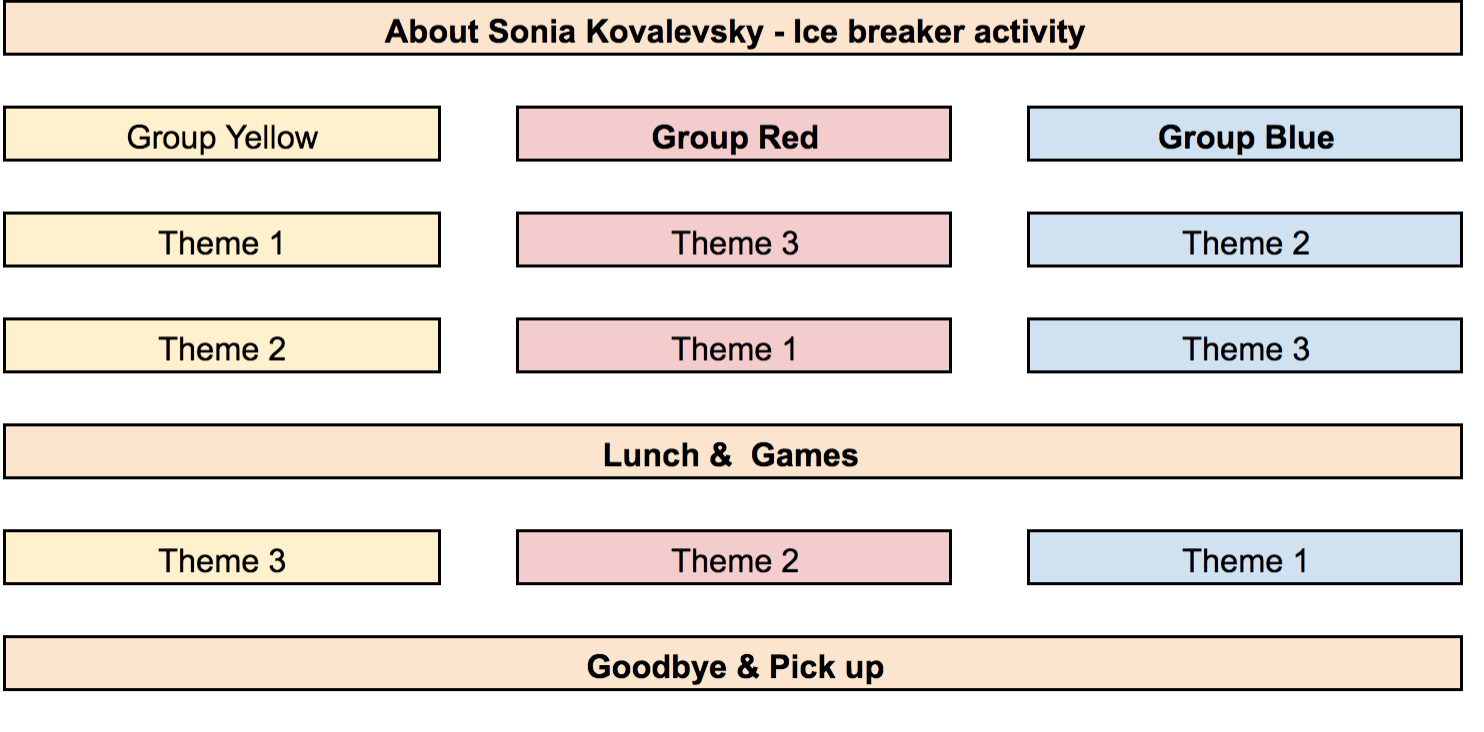}
\end{center}
 
 After a few years of modifying the schedule, it seems that sessions of 45 minutes work best, with 5 to 10 minutes breaks in between, and an hour lunch break (typically pizza!). We've usually run the program on Saturday's from 10:30 a.m. to 3 p.m.,  this is to accommodate families coming from outer Chicago but in a smaller city, the program could be longer.  
  \newpage
      \noindent {\bf Before the event}. There are five main tasks to take care of before the event, which can be done as little as a month in advance. In order of priority these are:
      
      \begin{itemize}
      \vspace{3mm}
      \item[(I)] {\it Set the date and book rooms}. 
      \end{itemize}
      
      The most successful days will be Saturdays which don't coincide with holidays or school activities. Contacting a few parents beforehand to check school's calendars has shown to be the most efficient way of choosing the date -- once this is done the rooms for the event should be booked. At UIC, we have ran these events in May and  November.

             \begin{itemize}

      \vspace{3mm}

      \item[(II)] {\it  Find volunteers}. 
            \vspace{3mm}

            \end{itemize}

It is useful to start looking for volunteers as early as possible and, if volunteers appear first, their schedule can be taken into account when setting the date. The easiest approach is to reach out to the graduate and undergraduate communities within the Mathematics and Physics departments -- at UIC, we make an effort to include both departments. Volunteers are usually female students, and if male students want to help, they are assigned non-teaching tasks, trying to maximise the opportunities for the young girls to find role models amongst the volunteers. 

It should be noted that too many volunteers can make the participants shy and the classes less interactive. We have found that 2 to 3 volunteers per room is ideal. Finally, universities typically require background checks for anyone working with minors, and the details of those present during the event must be given to the department officers several weeks in advance. 
        
               \begin{itemize}

      \vspace{3mm}
      \item[(III)] {\it  Advertising the event}.
      \vspace{3mm}
            \end{itemize}
A theme for the Sonia day should be chosen after setting the date for the event, in order to begin the advertisement. For this, a poster should be made, as well as a website with the relevant information  including date, place, schedule, parking directions and poster, sign up form (e.g. google form). Emails to departmental members asking to promote the event have always been very helpful, and contacting the school district offices for them to send an announcement to all schools in the area appears to be the best way to reach a wider range of participants.

      \begin{itemize}
         \vspace{3mm}
   \item[(IV)]{\it  Secure funding}.
            \vspace{3mm}

            \end{itemize}

 A SK day can be ran with very little funding, most of it being needed for providing lunch -- pizzas delivered to the event can make life easier, leading to about \$8 a person. If more funding were available (e.g. through an NSF  grant,  start up funds, or the Department's outreach program), it could be dedicated to purchasing breakfast treats,  pens, small notebooks and stickers with a small  poster   for the cover. Moreover, giving girls T-Shirts with the logo of the SK Day has made participants very happy, and allows for volunteers to be recognised. These extra things can add  \$10 per participant/volunteer.

        \begin{itemize}

            \vspace{3mm}
   \item[(V)] {\it  Material for the event}. 
            \vspace{3mm}
         \end{itemize}
            The SK organisation becomes much easier if as much as possible is  done some weeks in advance, and this includes taking care of: 

\begin{itemize}
\item[$\star$] Choosing the themes and preparing the notes for volunteers and students. 
\item[$\star$] Event folders with notes, notebook, pens and a few printouts about the AWM   and the hosting university's programs. 
\item[$\star$] Leaving survey for students.
\item[$\star$] Background checks.
\item[$\star$] Image release forms for parents. 
\item[$\star$] Posters and flyers. 
\item[$\star$] T-shirts for event. 
\item[$\star$] Arrival sign up list of students.
 \item[$\star$] Lunch time entertainment volunteer. 
  \item[$\star$] Room and building booking/arrangements. 

\end{itemize}
\columnbreak
    \noindent {\bf The themes}. The Sonia Kovalevsky days usually are organized with one overall theme, and three lectures surrounding these theme. Over the years we have found that mixing mathematics with real life problems or geometric ideas is a great way to get students interested.     
    We have now run the following themes:

    \begin{itemize}
    \item Games on Surfaces:
     \begin{enumerate}
    \item Billiards
    
    \item Non-orientable surfaces
    \item Sphere packing 
    
    \end{enumerate}
      \item Games of Chance:
     \begin{enumerate}
    \item Non-transitive Dice
    
    \item Sums of Dice
    \item The Monty Hall Problem
    
    \end{enumerate}  
    \item Knots and Graphs:
     \begin{enumerate}
    \item Graph Colouring 
    
    \item Mathematical knots
    \item Bridges of K\"onigsberg 
    
    \end{enumerate}
      \item Mathematics in the sea:
     \begin{enumerate}
    \item Seaweed Tangles
    
    \item Fractal Coastlines
    \item Geometry of Seashells
    
    \end{enumerate}
      \item Mathematics and Magic:
     \begin{enumerate}
    \item Flexagons
    
    \item Card Tricks
    \item Magic Squares
    
    \end{enumerate}
    \end{itemize}
    Each of these themes should be prepared with a set of notes including overall goals for the hour, additional bibliography on the subject, and a list of 5-10 problems of different difficulties for the students, particularly to account for the age range.  Material for the above subjects 
    and the manner it was presented to students and volunteers can be seen in \href{https://schapos.people.uic.edu/Outreach.html}{https://schapos.people.uic.edu/Outreach.html}. 
    
    For some of the classes we gave the students some objects they could take home with them -- e.g.~for learning about knots and dice we gave them ropes and  non-standard dice (made by us). Once the notes for the chosen themes are prepared, it is useful to have an organizational meeting with the volunteers prior to the event. 

\columnbreak  \noindent {\bf During the Event}.
When the SK day arrives, it is useful to have the volunteers come an hour earlier to help set up the main room: displaying the sign up sheet (attendance is usually 50\%-70\%), the photo release forms to be signed by parents, the T-Shirts and the activity folders. The main room is used for the presentation on Sonia Kovalevsky's life, the ice breaker, the lunch break and final participant's pick up. 

During the whole day it is useful for the main organizers to go around the different rooms, taking photos, helping encourage the  girls interact with each other, and making sure the schedule is being followed and to get lunch set up. At lunch time,  someone entertaining the participants  can be a highlight -- at UIC we have had Lou Kauffman and his students do magic tricks. 
 At the end of the day, the participants are asked to complete a short anonymous questionnaire about their experiences of the event (an example is given in the website), to identify any highlights or weaknesses that might be improved in future events. These surveys are then used to make an {\it Event Report}. 
 
   \noindent {\bf Concluding remarks}.
 The participants have been universally happy with the event, with 100\% positive replies on all aspects of the program every year. Moreover, they have always given a strong indication that they would return  to future events and encourage others to participate. Moreover, by mixing volunteers from Mathematics and Physics, many new friendships have been made. To thank the volunteers, we usually take them for lunch some weeks after the event, to hear about their thoughts on the event and give any career advice they may need. 
 
\vspace{2mm} \noindent {\bf Acknowledgements.}~The authors thank the Simons Center for Geometry and Physics for hospitality and support. LPS is supported by the Humboldt Foundation and NSF grants  DMS 1509693 \& CAREER DMS 1749013.
\vspace{2mm} 
 
\noindent {\it Email}: schapos@uic.edu; unwin@uic.edu.\\
 \small{University of Illinois at Chicago, IL.    \\Simons Center for Geometry  and Physics, NY.}

%
%
%
%
%

    \end{multicols}


\end{document}